\documentclass{article}

\usepackage{PRIMEarxiv}

\usepackage[utf8]{inputenc} 
\usepackage[T1]{fontenc}    
\usepackage{hyperref}       
\usepackage{url}            
\usepackage{booktabs}       
\usepackage{amsfonts}       
\usepackage{nicefrac}       
\usepackage{microtype}      
\usepackage{lipsum}
\usepackage{fancyhdr}       
\usepackage{graphicx}       
\graphicspath{{media/}}     

\usepackage{float}
\floatstyle{boxed} 
\restylefloat{figure}
\usepackage{graphicx}
\usepackage{caption}
\usepackage{subcaption}
\usepackage{url}

\usepackage{amsmath}

\usepackage{etoolbox}

\usepackage{array}

\title{Reduction of Two-Dimensional Data for Speeding Up Convex Hull Computation
\thanks{}
\textbf{}
}

\author{
  Debashis Mukherjee \\
  Associate Professor, Computer Science \& Engineering \\
  Budge Budge Institute of Technology \\
  Kolkata, India\\
  \texttt{\{debashismukherjee20\}\@gmail.com} \\
}

\begin{document}
\maketitle

\begin{abstract}
An incremental approach for computation of convex hull for data points in two-dimensions is presented. The algorithm is not output-sensitive and costs a time that is linear in the size of data points at input. Graham's scan is applied only on a subset of the data points, represented at the extremal of the dataset. 
Points are classified for extremal, in proportion with the modular distance, about an imaginary point interior to the region bounded by convex hull of the dataset assumed for origin or center in polar coordinate. A subset of the data is arrived by terminating at until an event of no change in maximal points is observed per bin, for iteratively and exponentially decreasing intervals.
\end{abstract}


\section{Introduction}
\label{sec1}

Computation of convex hull of planar datasets is well studied in the literature of computational geometry ~\cite{Jarvis, Graham, Barber, Chantm, Liu, Gomes}. 
A convex hull of a dataset is an ordering defined on the largest subset based on the geometry represented by the set.
In two-dimensional dataset, the ordering leads to a closed polygon, where every line segment with any two vertices from the set, is contained entirely in the region bounded by the polygon. Algorithms for construction of convex hull, bears a correspondence with the algorithms for sorting.

An algorithm for computation of convex hull is said to be output-sensitive, based on whether the complexity of computation depends on any structural attribute specific to the dataset at input other than its size.
A consistent traversal (e.g. a turn in a counter-clockwise (CCW) sense) of the data points in a set, is often considered in computation of convex hull.
An earliest output-sensitive algorithm by Jarvis et. al. ~\cite{Jarvis}, known as \textit{gift wrapping}, computes convex hull in $O(nh)$, for dataset of size $n$, and due to $h$ points on the hull. The algorithm called Jarvis march, starts from a point that is lowest in one coordinate, and then selects the next point on the hull, with a consistent turn, and corresponds to \textit{selection} sort.

Graham's scan is an earliest algorithm, that uses an explicit sort on the dataset, and then a linear time scan ~\cite{Graham}. The time complexity of $O(n \log n)$ of the \textit{mergesort}, or \textit{quicksort} step on average, bounds the running time of the algorithm. The algorithm is not output-sensitive, and also determines a lower bound on the complexity of computation of convex hull. Divide-and-conquer based strategy, splits dataset in pairs about pivots, and merges convex hulls of subsets in linear time, recursively, and resembles that of a \textit{quicksort}. The asymptotic average complexity of $O(n \log n)$, with randomly chosen pivot points in divide-and-conquer approaches, is improved further, in \textit{QuickHull} algorithm, by excluding data points that are interior to intermediate polygons from computation ~\cite{Barber}. Chan's algorithm, improves the complexity, of Graham's scan, to $O(n \log h)$, by applying the algorithm $\frac{n}{h}$ times, on smaller subsets of size $h$, and merging all subsets in linear time ~\cite{Chantm}. Several existing algorithms use incremental insertion of data points, resembling that of approaches of \textit{insertion} sort.
Improvement in speed by extending and inflating an intermediate hull incrementally about boundary points have also been presented by Liu et. al. ~\cite{Liu} and Gomes et. al ~\cite{Gomes}.



Table \ref{table:1} summarizes the algorithms for construction of 2D convex hull and the average case of complexity, that are available in the literature.

\begin{table*}[h!]
\centering
\begin{tabular}{| c | c | c |} 
 \hline
 Algorithm & Expected Time Complexity  & Reference \\ 
 \hline
 Gift Wrapping & $O(nh)$ & (Chand and Kapur, 1970) ~\cite{Chand}\\ 
 \hline
 Graham Scan & $O(n\log{n})$ & (Graham, 1972) ~\cite{Graham}\\
 \hline
 Jarvis March & $O(nh)$ & (Jarvis, 1973) ~\cite{Jarvis}\\
 \hline
 QuickHull & $O(n\log{n})$ & (Barber,Dobkin,Hannu 1996)  ~\cite{Barber} \\
 \hline
 Divide \& Conquer & $O(n\log{n})$ & (Preparata and Hong, 1977)  ~\cite{Preparata}\\
 \hline
 Monotone Chain & $O(n\log{n})$ & (Andrew, 1979)  ~\cite{Andrew}\\
 \hline
 Incremental & $O(n\log{n})$ & (Kallay, 1984)  ~\cite{Kallay}\\ 
 \hline
 Marriage before Conquest & $O(n\log{h})$ & (Kirkpatrick and Seidel, 1986)  ~\cite{Kirkpatrick}\\ 
 \hline
 Chan’s algorithm & $O(n\log{h})$ & (Chan, 1996)  ~\cite{Chantm}\\ 
 \hline
 Ordered hull & $O(n\log{h})$ & (Liu and Chen, 2007)  ~\cite{Liu}\\ 
 \hline
\end{tabular}
\caption{Summary of the algorithms for 2D convex hull construction.}
\label{table:1}
\end{table*}

Recent work, on convex hull computations, considers ways to reduce the set, to a subset of the data that is of interest. Skala et. al. ~\cite{Skala} also uses polar coordinates, to partition data points in polar sectors, about an origin computed using 10\% of the dataset, and then computes cluster of extremal points, and discards data points contained in a rectangular region. Cadenas et. al. ~\cite{Cadenas} uses bins in rectangular coordinates, to partition data points, about top and bottom boundaries, and then uses them to compute a subset representing a \textit{fence}, which is being output after the preprocessing steps for convex hull computation.

In the present work, we consider data points in two-dimensions only, and therefore, we use a circle for that of the sphere in our representation. The data points are ordered by a distance $r$ about the origin, at a polar angle $\theta$, for polar coordinates of $(r,\theta)$. The centroid, or the mean of the cartesian coordinates of the data points, is taken as the origin in our computation. The main contribution of the present work in comparison to the earlier works specifically by Skala et. al. and Cadenas et. al., 
is an way to arrive at a smaller set of extremal points precisely equivalent to contour of the geometry represented by the dataset.


The paper is organized as follows. At first, we discuss the related work and the background of our work. The main focus of the article is discussed next, including, definitions of basic concepts and terminologies, used in the rest of the paper. Subsequently, the algorithm addressing the problem in the title, is presented, and experimental results are discussed. The paper is concluded finally at the end.

\subsection{Related Work}
\label{sub_sec1}

Similar to that of Skala et. al., we also use polar sector, and consider them as bins, for data points  ~\cite{Skala}, and choose point that is farthest from the origin per bin, and connect them in order of polar angle. Skala et. al. uses, a line segment joining the farthest points between two consecutive sectors, to partition, the extremal data points in the interval between the pair of sectors. The newly identified set of the data points although add to a cluster, however any further reduction of the set, is not considered through the approach proposed by Skala et. al.  Instead of using the line segment, we use the angle subtended by the line at a farthest point with the origin of the polar coordinate, called horizon angle, to partition the data points at the boundary of the circle. We are also able to reduce the set of the clustered points, further using relation based on the horizon angle.

Similar to that of Cadenas et. al.  ~\cite{Cadenas} we also use the boundary points to lay a fence, and remove data point represented by certain bin, based on its left and right neighbors, to preserve a property of convexity in the fence. Instead of rebinning, proposed by Cadenas et. al, we introduce additional data points on the fence from that localized by the bins, using horizon angle. We also introduce data points, with the largest and smallest of each of the Cartesian coordinates on the fence, in order of the polar angles, for points corresponding to left, right, top and bottom boundaries and the beginning and end of the points in these on the convex hull. The data points included in the fence, finally in our case, is very close to the number   of the data points on the convex hull, and the total cost of computation of all our steps is $O(n)$.

Total order heuristic-based convex hull (TORCH) algorithm, proposed by Gomes et. al. ~\cite{Gomes}, computes convex hull for data points on the Cartesian plane, incrementally by inflating an intermediate hull obtained using ordering of the vertices near the corner and on the boundary. The data points at the boundary of the convex hull, computable in $O(n)$, is merged with the data points computed during the main course of the algorithm in both the current approach, and in the approach used in TORCH.  The TORCH algorithm, depends on sorting of the data points on a Cartesian basis at first, which limits it to $O(n\log{n})$, however a significant improvement in the running time over Quickhull, was claimed for large dataset, due to a use of efficient heuristics steps towards a shape at the goal of the hull. Several recent work are a variant of the approach towards incrementing an initial polygon to the target convex hull on sorted planar dataset ~\cite{Ferrada,Alshamrani,Mei}.

\section{Present Work}
\label{_sec2}

The focus of the current article, is to present an algorithm, that can be used for speeding up of convex hull computation for data points on a plane. An arbitrary plane in the three-dimensional Euclidean space, is assumed to be transformed, to one of the principal planes.  The data corresponding to the set of the input, is therefore, considered to be given on the two-dimensional Euclidean space, with specific two basis vectors $X$, and  $Y$ only.

The algorithm presented in this paper, can be used to reduce the set of input data points, to a subset, which is very close to the number of points on the convex hull. We define the following terminologies, which are necessary for presentation of the algorithm.

\begin{figure}
    \centering
    \begin{subfigure}[b]{1\textwidth}
        \includegraphics[width=\textwidth]{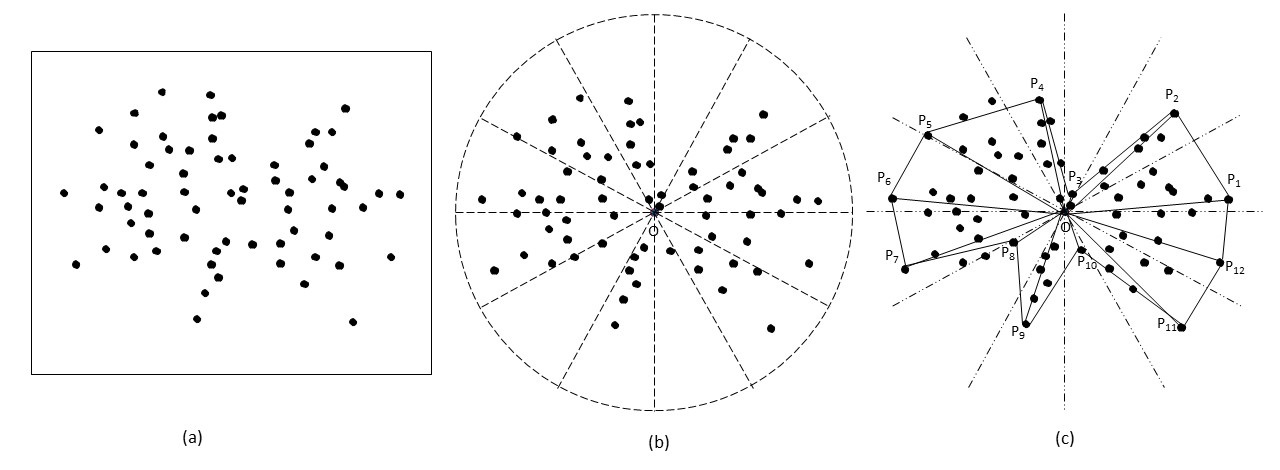}
    \end{subfigure}
    ~ 
    \begin{subfigure}[b]{1\textwidth}
        \includegraphics[width=\textwidth]{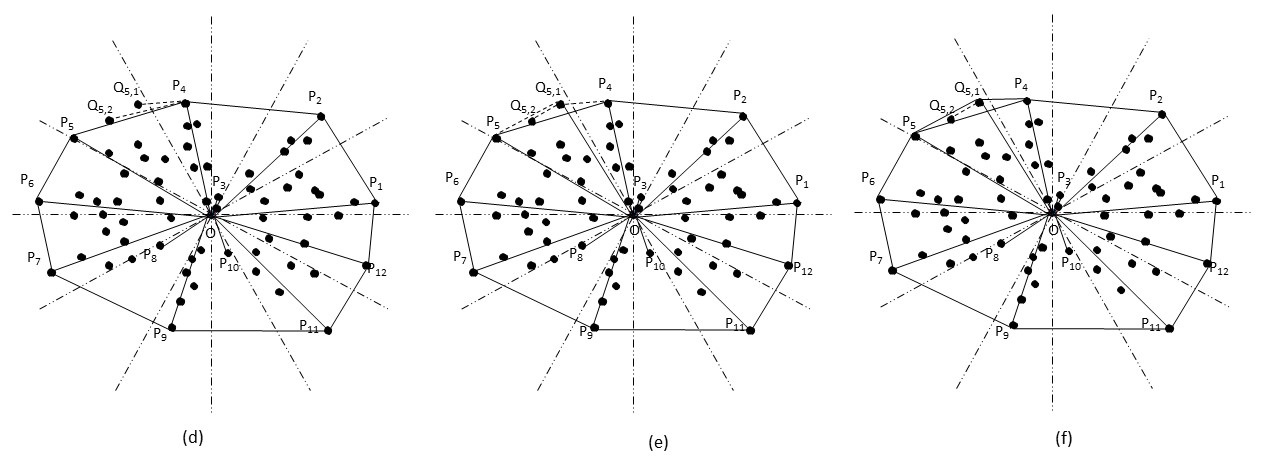}
    \end{subfigure}
    ~ 
    \begin{subfigure}[b]{1\textwidth}
        \includegraphics[width=\textwidth]{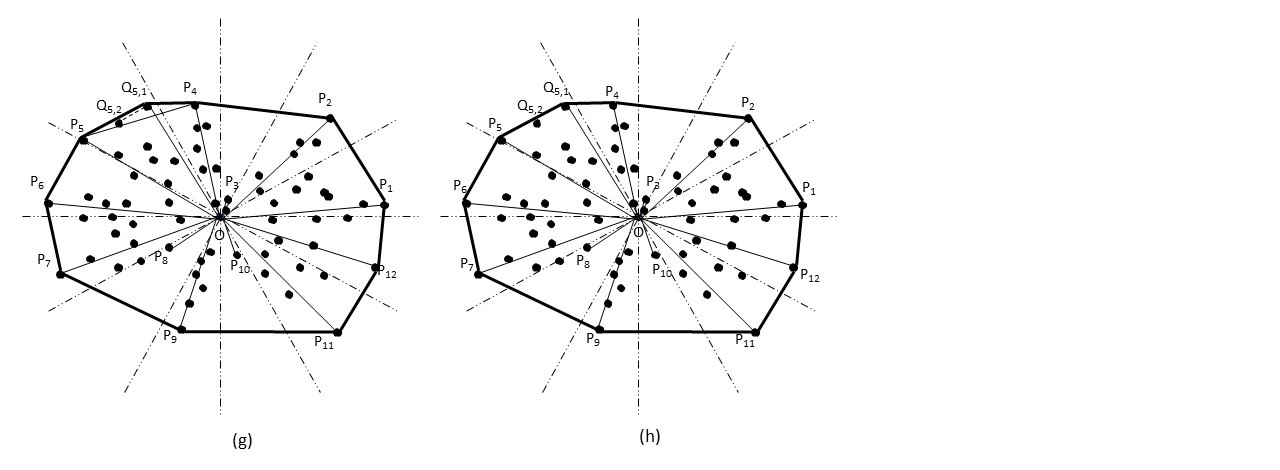}
    \end{subfigure}
    \caption{Construction of 2D convex hull. a) Input data points, b) Binning of data points, c) Fence after binning, and boundary computation, d) Convexity preservation of fence, and beginning of horizon computation, e) Identification of horizon set (near bin 4 and 5), and computation of horizon points, f) Completion of horizon computation, g) Fence after horizon computation (in thick line), h) Completed fence for convex hull computation on termination including, elements of maximal bin point set, boundary point, horizon points. }\label{res5_3dfig_1}
\end{figure}

\subsection{Terminologies}
\label{sub1_sec2}
\begin{itemize}
    \item \textbf{Center} : A center is a hypothetical point, corresponding to the set of input data points. The Cartesian coordinate of the center is computed using the arithmetic mean of the Cartesian coordinates of the input data points. The coordinate $(x,y)$ of of the center $O$ is computed using, $O.x=$$\frac{1}{n}$$(P_{1}.x+...+P_{n}.x)$, and $O.y=$$\frac{1}{n}$$(P_{1}.y+...+P_{n}.y)$. We translate the set of data points, such that the center becomes the origin, i.e,  $O.x=0$, $O.y=0$.
    
    \item \textbf{Polar coordinate} : The polar coordinate of a data point, is computed from its cartesian coordinate, and the center. The polar coordinate $(r,\theta)$ of a point $P_{i}$, is computed using $r.i=$$\sqrt{{P_{i}.x}^{2}+{P_{i}.y}^{2}}$, and $\theta.i={\tan}^{-1}(\frac{{P_{i}.y}}{{P_{i}.x}})$.
    
    \item \textbf{Farthest point} : The farthest point is a data point with largest radial distance about the center. The polar coordinate $(r,\theta)$  of farthest point $P_i$, is such that $r.i$, equals $\max(r.0, ..., r.n)$. The farthest point is a point on the fence and on the convex hull.
    
    \item \textbf{Boundary point} : A boundary point, is a point, with either a maximum or a minimum cartesian coordinate, from the set of the input data points. Boundary points are points on the convex hull and on the fence. Point $P_{1}$  and point $P_{6}$ , are boundary points in Figure 1, based on maximum and minimum $X$ coordinate respectively.

     \item \textbf{Polar bin} : A polar bin, is a fractional part of a circle, based on an angle lesser than $2\pi$  subtended at the center by a pair of radii of a circle, for a circle with radius greater than the farthest point. The angle $\phi$  subtended by the radii at the center is called the \textit{bin interval}. The $i$'th polar bin $B_{i}$, corresponds to the fractional part of the circle, from polar angle $\phi(i-1)$ to $\phi$$i$. The input point $P_{1}$ in Figure 1.c, is contained in polar bin $B_{1}$.
     
     \item \textbf{Maximal bin point set} : Maximal bin point set, is a subset of data points, contained in a polar bin, with the maximum radial distance about the center, among all data points contained in the bin. The size of the maximal bin point set of polar bin $B_{1}$ in Figure 1.c) is one, and input point $P_{1}$ is the element of the set. 
    
    \item \textbf{Horizon angle} : Horizon angle in the angle subtended at an element of a maximal bin point set, with any other point and the center. A horizon angle $ha(m,p)$ denotes the angle subtended at the element $m$ of the maximal bin point set, with point $p$ and the center. The point about which the angle is subtended is called the \textit{anchor} point of the horizon angle. The angle ${Q_{5,1}P_{4}O}$ subtended at the maximal bin point $P_{4}$  with point $Q_{5,1}$  and the center $O$, in Figure 1.d)  is the horizon angle of  the input data point $Q_{5,1}$  and is denoted by  $ha(P_{4},Q_{5.1})$.

    \item \textbf{Traversal order} : A traversal order, is the order of the polar angle in which the next point is visited. The default traversal order is considered in increasing order of polar angles, and is anticlockwise.
    
    \item \textbf{Horizon set} : Horizon set $hs(i)$ is defined as a subset of the set of points from the interval between the polar angles subtended by maximal data point represented by two consecutive bins indexed $i$ and its next. The set is defined with reference to the horizon angle $ha(m1,m2)$  of any one of the maximal data points $m1$, with the other $m2$. An element of the horizon set, is a point, that subtends a larger horizon angle, at the maximal data point, than horizon angle subtended by the other maximal data point, or $p \in hs(i)$, if  $ha(m1,p) \ge ha(m1,m2)$.  The horizon set corresponding to bin indexed 4 ($B_{4}$) with its next, is $\{Q_{5,1}, Q_{5,2}\}$ in Figure 1.d), since we have  $ha(P_{4},Q_{5,1})$, $ha(P_{4},Q_{5,2})$, for elements $Q_{5,1}$, and $Q_{5,2}$ respectively.
    
   \item \textbf{Horizon point} : A horizon point is an element of a horizon set, with the largest horizon angle subtended at its \textit{anchor} point. A point identified as a horizon point, is considered equivalent to the status of an element of the maximal bin point set, in subtending horizon angle, or can be considered as an \textit{anchor} point, for computation of its next horizon point. Point $Q_{5,1}$ is a horizon point in Figure 1.e), because for the horizon set $\{Q_{5,1}, Q_{5,2}\}$, we have $ha(P_{4},Q_{5,1}) > ha(P_{4},Q_{5,2})$. The point $Q_{5,1}$, is therefore considered as an anchor point, for computation of any horizon point, at its next. But, the horizon angle to the end of the interval at the new anchor point $Q_{5,1}$ is $ha(Q_{5,1}, P_{5})$, becomes larger than $ha(Q_{5,1}, Q_{5,2})$, i.e, the horizon angle with the only other remaining element from the horizon set, in this case, So, no new horizon point is found at the next of the horizon point $Q_{5,1}$ , and the sequence of the horizon points found is $(Q_{5,1})$  only, in this case.    
    
  \item \textbf{Fence point list} : Fence point list, is a list of points, which are either elements of maximal bin point sets, or boundary points, or horizon points, ordered on the polar angle, in a traversal order. Every polar bin in a sequence of the 12 bins in Figure 1.c) has a maximal bin point set of size one, and we have a fence point list of the same length, i.e,  $(P_{1}, P_{2}, ..., P_{12})$.    
  
   \item \textbf{Convexity property of fence point list} : Convexity property is defined to preserve a convex structure in the sequence of the data points included in the fence point list. A point that fails to preserve the convexity property, is removed from the fence point list. A point fails to preserve the convexity property, if the line segment formed by joining the points, previous and next of it, finds the point, and center at the same side of the line. Maximal points represented by bins 3, 8, and 10, in Figure 1.c, fails to preserve the convexity property, therefore these points are removed from the fence point list.

    \item \textbf{Contour curve} : 
    A contour curve is a closed and continuous curve, containing \textit{maximal bin point}s and \textit{boundary point}s of the dataset.




\end{itemize}

\subsection{Algorithm: Input Reduction for 2D Convex Hull}
\label{sub2_sec2}

The algorithm for computation of the fence point list, which is actually a subset of the input, is presented in the following. Computational cost of a construction of a convex hull using traditional algorithms, would be much lower with the reduced point set, than that with the original input data set, depending on the input.

The steps used in the preprocessing for speeding up of convex hull computations are: i) Binning, ii) Boundary computations, iii) Fencing, iv) Horizon computations, 

\textbf{Step 1 (Binning)}:  The internal center point of the data and the polar coordinates are computed from the Cartesian coordinates of the input at the beginning of this step. The binning step assumes a bin interval, and creates the required number of polar bins. A data point is assigned to a bin, if its polar angle is in the interval covered by the bin. 
The radial distance of the maximal point of the bin is compared with the radial distance of the data point, and the \textit{maximal bin point set} for the bin is updated.
Data points are assigned to the bins, and maximal bin point sets are computed for each polar bin, by the end of the binning, in $O(n)$ time.
The starting of the \textit{bin} indexed \texttt{0} ($B_{0}$) is ensured to be either a \textit{boundary point} or \textit{farthest point}.


\textbf{Step 2 (Boundary computations)}:  The boundary data points of the input, are included in the convex hull. These points are computed in this step, and inserted in the fence point list in order of the polar angles. This step takes $O(n)$ for completion and can be interleaved with binning step, 

\textbf{Step 3 (Fencing)}: Fencing is an intermediate step, that arranges a fence point list, consisting of data points of maximal bin point sets, boundary points, and horizon points, in an order of the polar angles.  Fencing is performed at the end of each step, namely binning, boundary computation, and horizon computation, subsequent to the identification of new points, that may be considered for an inclusion on the convex hull. Fencing is interleaved with the later steps, and takes $O(n)$ for completion.

\textbf{Step 4 (Horizon computations)}: Horizon computation step, includes computation of horizon sets at first, and then using the set to compute a sequence of horizon points, in the interval of the polar angle represented by the horizon set. Computation of horizon set requires one iteration of the points in the interval between a pair of bins, and can be skipped for bins, which represents points that, fails to preserve convexity property in the fence. Thus, computation of all horizon sets takes at the most $O(n)$  time for completion, and is the major part of the step. The size of the horizon sets is usually very small on average if the bin intervals are within a medium size range. For an average size $s$  of a horizon set, the average cost of computation of a sequence of horizon points is $O(slogs)$, and for $t$  bins, the total cost for the remaining part of the step is at most $O(tslogs) \le O(n)$  as $s \ll \frac{n}{t}$. Thus, the step takes at the most $O(n)$  time for completion.



\textbf{Step 5 (Contour scanning)}: In this step we apply Graham's scan ~\cite{Graham}, on the dataset representing the \textit{contour}, in order of increasing polar angles, to compute a convex hull. The hull thus computed of the \textit{contour}, is also the convex hull for the input dataset. The step takes $O(m)$ time, for dataset of size $m$ on the contour. 

The incremental steps used in our algorithm, for computation of convex hull, by reduction of datasets on a plane based on \textit{horizon computation}, could be summarized as follows:

\begin{enumerate}
\item Perform \textit{binning} with a bin interval of $10$ \textit{degree}s, i.e. $\frac{\pi}{18}$ \textit{radian}, or lower.
\item Perform \textit{fencing} by merging maximal bin points, with boundary points, in order of polar angle
\item Perform \textit{horizon computation}, and obtain reduced set of extremal points.
\item Apply traditional algorithm for construction of convex hull on the reduced dataset.
\end{enumerate}

The incremental steps used in our algorithm for construction of the convex hull, based on \textit{contour scanning} could be summarized as follows:

\begin{enumerate}
\item Perform \textit{binning} with an initial bin interval of $1$ \textit{degree}, or $\frac{\pi}{180}$ \textit{radian}.
\item Perform \textit{binning} in repetition with half of the previous bin interval, and stop at interval step, where, lowering, has no change in corresponding \textit{maximal bin point set}.
\item Apply Graham's algorithm, for the set of points ordered on the contour.
\end{enumerate}

  

\subsection{Experimental Results}
\label{sub3_sec2}


We have used \texttt{Python 3.8} in an environment under the \texttt{Anaconda} software for proof-of-concept validation of the algorithms presented in Section \ref{sub2_sec2}. The system included, an ordinary Lenovo (IdeaPad) Laptop computer, Windows 10 Home Edition OS, Intel(R) i3 CPU @2.30Hz, and 12.0 GB RAM. The \texttt{Python} code and the datasets used for the experimentation, are available at a public repository under GitHub ~\cite{GitHub}. Validations were possible in platforms for open source \texttt{Python} code, within \texttt{Jupyter Notebook} on \texttt{Anaconda} installation.

We have compared the running time in \texttt{Second}s for various datasets, with and without applying a reduction presented in \textit{horizon computation}, by \textit{Jarvis}'s algorithm, and with our \textit{contour scanning} algorithm. We have used datasets used in the TORCH literature ~\cite{Gomes}.  Our experimentation with the datasets is reported in Table \ref{table:2}. The specification for the datasets are shown in columns \texttt{1}, and \texttt{2}. The percentage of reduction in the size of the dataset actually used by our approach for \textit{horizon computation} are shown in column \texttt{3}. The time required for computation of the convex hull directly from the input dataset using \textit{Jarvis march} algorithm is shown coumn \texttt{4}.    The column \texttt{5} reports an ordered pair of time observed, in computation of the hull, indirectly by \textit{Jarvis march} on the reduced subset of the input using our \textit{horizon computation} scheme, and by our \textit{contour scanning} algorithm only, respectively. 
The \textit{bin interval} used for \textit{horizon computation} was \texttt{2} degrees, whereas \textit{bin interval} of \texttt{1} degree, was arrived for \textit{contour scanning}.

\begin{figure}[!h]
    \centering
    \begin{subfigure}[b]{1\textwidth}
        \begin{center}\includegraphics[width=100mm, scale=0.75]{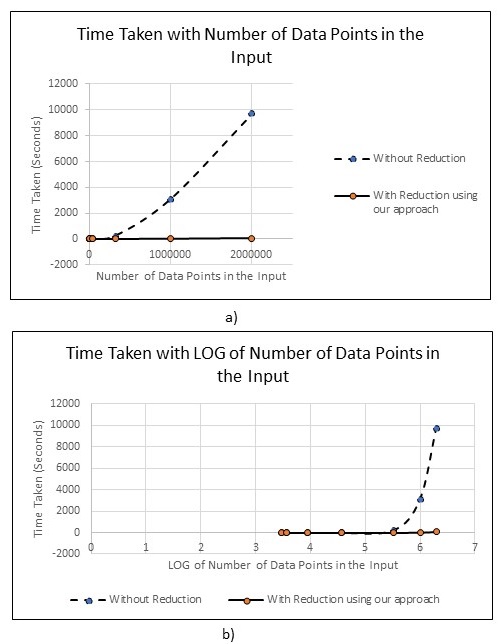}\end{center} 
    \end{subfigure}
    \caption{Summary of experimental observations on improvement in time taken in the construction using the reduction scheme. a) Time taken with size of input dataset, b) Time taken with $\log_{10}$ of the size of input dataset }\label{_plotfig_3}
\end{figure}

\begin{figure}[!h]
    \centering
    \begin{subfigure}[b]{1\textwidth}
        \includegraphics[width=\textwidth]{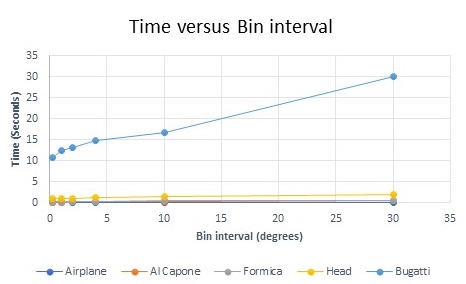}
    \end{subfigure}
    \caption{Effect of \textit{bin interval} on time of computation. }\label{_plotfig_4}
\end{figure}

The plots in Figure \ref{_plotfig_3} summarizes the time taken in a computation, with a size of the input, for various datasets, corresponding to the results of the experimentation reported in Table \ref{table:2}. The time after reduction using \textit{horizon computation} is shown in \texttt{solid} line, and the time for the algorithm (\textit{Jarvis march}) without any reduction on the input, is shown in \texttt{dash} line, respectively.   The plot in Figure \ref{_plotfig_3}.a shows a steep jump in the gradient, or an exponential growth in the time taken, is clearer without reduction, whereas a  linear and very slow growth, indicated by much lower, and consistent gradient is observed with our approach of \textit{horizon computation} and related reduction. A corresponding plot in scale of $\log_{10}$ for size of input dataset shown in Figure \ref{_plotfig_3}.b, depicts the significant growth observed beyond a size of $10^{5.51} \approx 320,119$ in a dataset in our case. The plot \ref{_plotfig_3}.b, when compared with \ref{_plotfig_3}.a, indicates a significant effect of the bound by the factor $\log_{10}(n)$, with growth of size $n$ of the dataset. A low value for \textit{bin interval} is observed to improve time for \textit{horizon computation}.
The computation time, for various \textit{bin interval}s on datasets are shown in Figure \ref{_plotfig_4}.

\begin{figure}[!h]
    \centering
    \begin{subfigure}[b]{0.4\textwidth}
        \includegraphics[width=\textwidth]{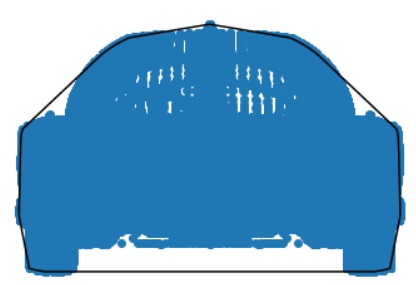}
        \caption{}
        \label{fig:gull}
    \end{subfigure}
    ~ 
    \begin{subfigure}[b]{0.4\textwidth}
        \includegraphics[width=\textwidth]{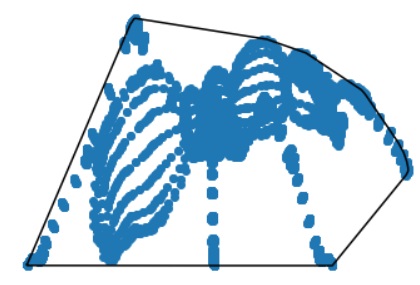}
        \caption{}
        \label{fig:tiger}
    \end{subfigure}
    ~ 
    \begin{subfigure}[b]{0.4\textwidth}
        \includegraphics[width=\textwidth]{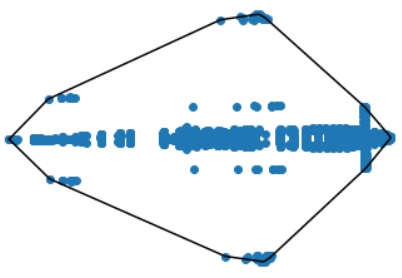}
        \caption{}
        \label{fig:mouse}
    \end{subfigure}
    \begin{subfigure}[b]{0.4\textwidth}
        \includegraphics[width=\textwidth]{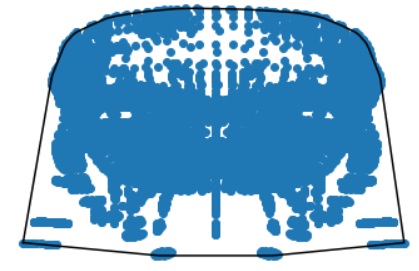}
        \caption{}
        \label{fig:gull}
    \end{subfigure}
    \begin{subfigure}[b]{0.4\textwidth}
        \includegraphics[width=\textwidth]{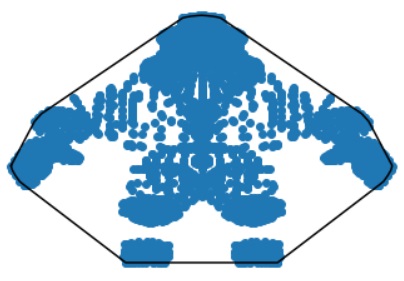}
        \caption{}
        \label{fig:mouse}
    \end{subfigure}
    \begin{subfigure}[b]{0.4\textwidth}
        \includegraphics[width=\textwidth]{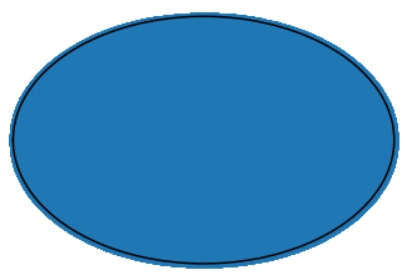}
        \caption{}
        \label{fig:gull}
    \end{subfigure}
    \caption{Convex hull construction in 2D for large regular datasets: a) Bugatti, b) Formica, c) Airplane, d) T800 Head, e) Al Capone, f) Circle 1. }\label{_exp1fig_4}
\end{figure}

A visualization of the datasets, reported in Table \ref{table:2}, and the computed convex hulls, are shown in Figure \ref{_exp1fig_4}.
The time in \texttt{second}s corresponds to a naive implementation of the algorithm in \texttt{Python} for a proof-of-concept only primarily.
An implementation leveraging vectorization and corresponding programming features for such large datasets could possibly reduce it further.

\begin{figure}[!h]
    \centering
    \begin{subfigure}[b]{0.4\textwidth}
        \includegraphics[width=\textwidth]{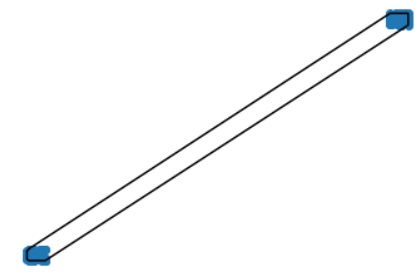}
        \caption{}
        \label{fig:gull}
    \end{subfigure}
    ~ 
    \begin{subfigure}[b]{0.4\textwidth}
        \includegraphics[width=\textwidth]{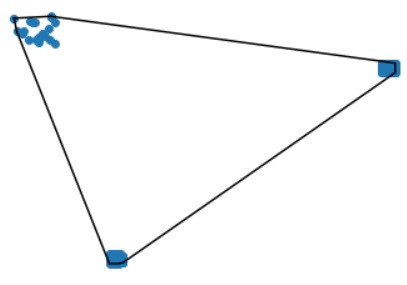}
        \caption{}
        \label{fig:tiger}
    \end{subfigure}
    ~ 
    \begin{subfigure}[b]{0.4\textwidth}
        \includegraphics[width=\textwidth]{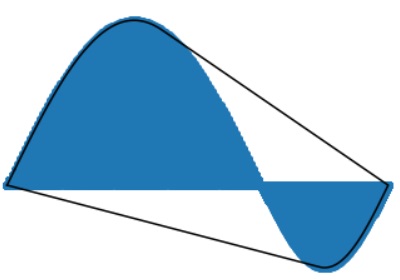}
        \caption{}
        \label{fig:mouse}
    \end{subfigure}
    \begin{subfigure}[b]{0.4\textwidth}
        \includegraphics[width=\textwidth]{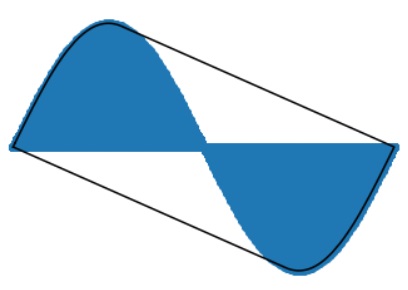}
        \caption{}
        \label{fig:gull}
    \end{subfigure}
    \caption{Convex hull construction in 2D for irregular distribution in datasets: a) Blobs, b) Clusters, c) Sinusoid 1, d) Sinusoid 2.  }\label{_exp1fig_5}
\end{figure}

Figure \ref{_exp1fig_5} visualizes a variety in the distribution of the datasets, used for validating the robustness of our algorithms, in another set of experimentation.
We have considered blobs, clusters, and irregularity in pattern in the data contained in the polar bins including cases for emptiness or size equals \texttt{0}.
The data contained in the sectors, about the \textit{center}, could be observed to be discontinuous for the datasets of Figures \ref{_exp1fig_4}.a, \ref{_exp1fig_4}.b, and \ref{_exp1fig_4}.d.
Our algorithm, could compute the convex hull for these cases, with no exception from that of dataset in \ref{_exp1fig_4}.c, and datasets in Figure \ref{_plotfig_3}.

A sketch for a proof to verify the correctness of our algorithm is summarized in the \textit{appendix} Section.

\begin{table*}[!h]
\centering
\begin{tabular}{| >{\small}c | >{\small}c | >{\small}c |  >{\small}c | >{\small}c |} 
 \hline
  Point Set & Points & Reduc. $\%$ & Time (sec.) (Jarvis) &  Time (sec.) ([Hor., Con.])\\ 
 \hline
 Airplane & 2,919 & 98.97 & 0.716 & [0.145, 0.10] \\ 
 \hline
 Al Capone & 3,618 & 98.59 & 1.267 & [0.206, 0.12] \\
 \hline
 Formica & 8,718 & 98.76 & 2.196 & [0.356, 0.26] \\
 \hline
 T800 Head & 35,925 & 99.88 & 12.481 & [1.097, 0.93] \\
 \hline
 Bugatti & 320,119 & 99.95 & 219.802 & [12.557, 9.55] \\
 \hline
 Circle 1 & 1,000,000 & 99.96 & 3072.814 & [49.075, 29.59] \\
 \hline
 Circle 2 & 2,000,000 & 99.98 & 9684.644 & [110.626, 70.07] \\
 \hline
 \end{tabular}
\caption{Summary of the results of experimentation and characteristics of the dataset used.}
\label{table:2}
\end{table*}



\subsubsection{Comparison with the state-of-the-art of the incremental approaches}
The incremental approaches ~\cite{Gomes,Ferrada,Alshamrani,Mei} apply on a sorted dataset, and ensures the best-case running time of the quickhull algorithm.
A largest possible bounding polygon, with vertices from the pointset is used to reject points interior to early in the further computation. Mei et. al. ~\cite{Mei} uses quadrilaterals in steps to discard data points, whereas Ferrada et. al. ~\cite{Ferrada} uses an octagon for the polygon  in the computation.
The present approach advances the state-of-the-art by using a polygon with a large number of sides, which in the limit is a circle, resembling the shape close to the geometry of a contour from the cloud of the points. 
The present approach, then expands the contour, by preserving convexity among neighbouring vertices of the polygon.
The next neighbour for a vertex, becomes possible to be determined without depending on explicit sort, by binning points on polar angles about an internal point as origin, and by saving a point of maximal modulus in polar coordinates upon entry of every point in a bin.     
In order to be operable with any intervals for bins, and practically medium-sized predefined value, a next neighbour in the present scheme is determined through step named \textit{horizon computation} subsequently, in $O(n)$ computation, in supplementary.
Performance of the scheme is observed to improve on smaller intervals, and case for the best interval is met in a limit, from no change with the rebinning successively.
The step for preservation of convexity, on points that are ordered, is just the linear and the later part only of the state-of-the-art Graham's scan algorithm.

\begin{figure*}[tp]
\centering
\includegraphics[width=14cm,height=10cm,keepaspectratio]{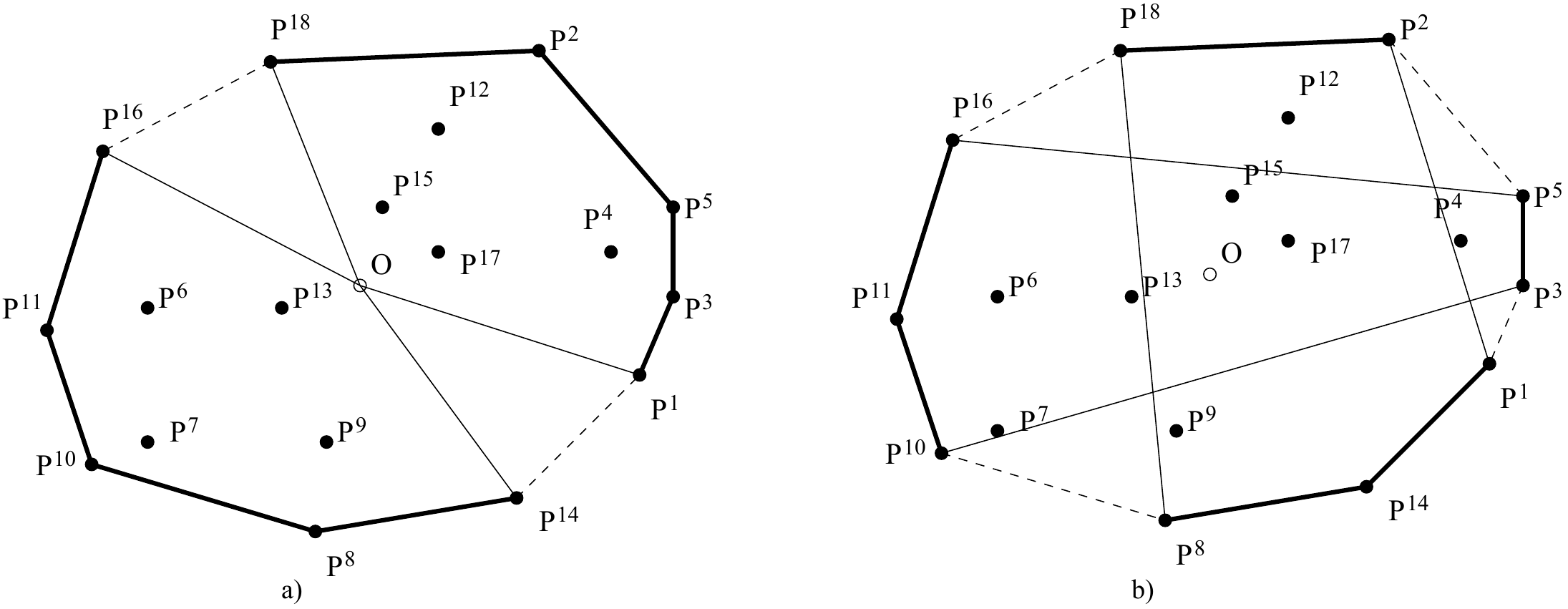} 
\caption{Representation based on \textit{Divide-and-Conquer} on the types of the \textit{divide} steps : a) Ordered on polar angle, b) Unordered or random }  
\label{_exp1fig_6}
\end{figure*}

\subsubsection{Comments on Time Complexity}

We have presented \textit{horizon computation}, and \textit{contour scanning} schemes, for computation of convex hull for datasets on a plane.
The \textit{binning} step is common for both, and costs $O(n)$, for input dataset of size $n$.
The cost of the steps \texttt{2} and \texttt{3}, together in \textit{horizon computation}, is $O(n)$ to arrive at a reduced set of size $n^{'}$.
The cost of the step \texttt{4}, for application of best known algorithm on the reduced dataset is $O(n^{'} \log n^{'})$.
Cost of construction, with reduction using \textit{horizon computation}, is $O(n)$ + $O(n)$ + $O(n^{'} \log n^{'})$ = $O(2n + n^{'} \log n^{'} )$.
The cost of the step \texttt{3}, in \textit{contour scanning}, is $O(m)$ for dataset of size $m$ on the \textit{contour}.
The cost due to the step \texttt{2}, depends on the iterations, required to arrive at the infinitesimal \textit{bin interval}, and could be only $O(n)$, with an appropriate initial interval.
Cost of construction, using \textit{contour scanning}, is $O(n)$ + $O(n)$ + $O(m)$ = $O(2n + m)$.
The cost of construction, using \textit{horizon computation}, is greater than \textit{contour scanning}, because, $n^{'} \ge m$.

The aforementioned algorithm for the construction can be formulated using a divide-and-conquer strategy.
A problem, of size $n$, can be divided into two \textit{subproblem}s of size $\frac{n}{2}$, with cost of \textit{divide} and \textit{combine}, denoted by $D(n)$, and $C(n)$, respectively.
The total cost of computation, can be expressed as, $T(n) = T(\frac{n}{2}) + T(\frac{n}{2}) + D(n) + C(n)$.
The divide step, can be specified, using the following two schemes:
\begin{enumerate} 
\item \textit{Ordered on Polar angle}: In this scheme, the polar angle for each data point in one subproblem, is always larger than the polar angle for each data point in the other subproblem. The data points in a subproblem, are taken from a distinct interval of the polar angles, which has no overlap, with the other subproblem.  
\item \textit{Unordered}: In this scheme, data points in the subproblems, are randomly selected from the original dataset.
\end{enumerate}

The \textit{divide} step, is considered at the end of the $binning$ step in our algorithm, and the cost of the \textit{divide} step $D(n)$ is $c_d$, where $c_d$ is a constant.
The precomputed \textit{center} of the original dataset at the input, is added as an additional data point, into the dataset of every subproblem. 
The \textit{combine} step, is considered after the subproblems are solved independently using our algorithm. 
In case of \textit{divide} using the \textit{ordered} scheme, the convex-hulls computed, are disjoint, as they are taken from distinct intervals of polar bins. 
The \textit{combine} operation $C(n)$, on the convex polygons at the output, therefore, applies only at the points on the boundary of the intervals of the polar angles in the corresponding polygons, and is possible in constant time.
Therefore, the cost of the \textit{combine} operation, for the \textit{ordered} case, is $c_o$, which is a constant.
In case of \textit{divide} using the \textit{unordered} scheme, the convex-hulls computed, may not be completely disjoint regions, as they may share intervals of polar bins.
The \textit{combine} operation, in such case, would require, recomputation of a convex-hull, using the set of data points on the convex-hulls of the subproblems.
Thus, the \textit{combine} operation $C(n)$, in such case, would require a cost of, $O(h_1 + h_2)$, where $h_1$ and $h_2$ are the number of points on the convex-hulls of subproblems, 1 and 2 respectively, $h_1$,$h_2$ $\ll$ $n$ is assumed on average, and $h=h_1+h_2$. 

The sum of the costs for \textit{ordered} \textit{divide}, is therefore, $T(n) = T(\frac{n}{2}) + T(\frac{n}{2}) + D(n) + C(n)$ = $2T(\frac{n}{2}) + c_d + c_o$ = $O(n)$,
and the sum of the costs for \textit{unordered} \textit{divide}, is therefore, $T(n) = T(\frac{n}{2}) + T(\frac{n}{2}) + D(n) + C(n)$ = $2T(\frac{n}{2})+ c_d + O(h)$ $\approx$ $O(n)$.

Ordered and unordered divide and combine scheme is shown in Figure \ref{_exp1fig_6}.
Figure \ref{_exp1fig_6}.a and  \ref{_exp1fig_6}.b shows the \textit{divide} and \textit{combine} steps for an ordered and unordered selection of data points, respectively.
The segments in \texttt{solid} line indicates the computation possible to be reused from the \textit{conquer} steps on subproblems in a divide-and-conquer scheme, and the segments in \texttt{dash} line correspond to those introduced in a \textit{combine} step, subsequent to their  respective \textit{divide} steps. Figure \ref{_exp1fig_6}.b depicts an instance of higher cost in the \textit{combine} step, by showing larger number of \texttt{dash} lines computed in the \textit{combine} step, compared to a case of \textit{divide} with an ordering on the polar angles of the dataset as shown in Figure \ref{_exp1fig_6}.a


\section{Conclusions}
\label{_sec3}

We have presented an incremental approach for construction of convex hull on a plane. The sorting of the dataset in our case is realized using a map onto a sphere. Thereby, our algorithm can compute a convex hull in $O(\alpha \times n)$ time, for constant $\alpha \le 2$. The present algorithm incorporates state-of-the-art Graham's scan into reduction of dataset at input for construction of convex hull related with a recent work ~\cite{Skala, Cadenas}. We have also merged boundary points with maximal bin points, for a contour determining the convex hull, incrementally ~\cite{Liu, Gomes}, early in the construction. Our algorithm is not output sensitive, and the steps of binning and contour interpolation are parallelizable. The transformation as that of a planar case may apply for higher dimensions and could be considered in any future work. The speed and robustness of our algorithm has been experimentally validated on large and irregular datasets.

\bibliographystyle{unsrt}  
\bibliography{references}  

\section*{Appendix}
\label{_sec3}

\textbf{Proposition 1: Center is an imaginary point contained in the region bounded by the convex hull}:  In our case, the cartesian coordinate of the \textit{center} $O$, defined in Section \ref{sub1_sec2}, is an arithmetic mean of the cartesian coordinate of the dataset.
As, there exist, some $i$, $j$, $k$, $l$, such that $(P.x)_{i} \le O.x \le (P.x)_{j}$ holds, and $(P.y)_{k} \le O.y \le (P.y)_{l}$, therefore \textit{center} is an internal point.

\textbf{Proposition 2: A point on a convex hull is a maximal bin point for some bin}:  There exist an infinitesimal bin interval $\delta \theta$, at the origin or center, with the bin containing the point $Q(r,\theta)$.
In the limit, when $\delta \theta$ goes to \texttt{0}, all points in the bin, are included on the radial line connecting the center to the point.
Now, a point $P_i(r_i,\theta)$, will not be an extremal point, if $r_i < r$.
Point $Q$ will be maximally extremal, or a maximal bin point for the bin  and therefore on the convex hull, if $r=\max_{i \in n}(r_i)$.

\textbf{Proposition 3: Contour scanning includes every point on the convex hull}:  In \textit{contour scanning}, we arrive at infinitesimal \textit{bin interval} interatively, by further lowering of \textit{bin interval}, until no change in \textit{maximal bin point set} is observed.
In such case, every point on the \textit{contour} is a \textit{maximal bin point} for some bin, and by use of Proposition 2, a contour is a superset of the set of data points on a convex hull.

\textbf{Proposition 4: A point on a convex hull is a horizon point}:  For any two consecutive points $P_i(r_i,\theta_{i})$ and $P_{i+1}(r_{i+1},\theta_{i+1})$ on a convex hull, let us assume there exists a point $P_{j}(r_j,\theta_{j})$, such that $\theta_{i} \le \theta_{j} \le \theta_{i+1}$, and angle $P_{i+1}P_{i}O$ is smaller than angle $P_{j}P_{i} O$. In such case, for \textit{anchor point} $P_i$, the point $P_{j}$ is a \textit{horizon point} at the next of point $P_{i}$ on the convex hull, and not $P_{i+1}$. The initial assumption that $P_{i}$ and $P_{i+1}$ are consecutive points on the convex hull would then be false, or contradiction is reached.

\textbf{Proposition 5: Horizon computation includes every point on the convex hull}:  Horizon computation involves computation of a convex set, from the set of all \textit{horizon points}. Horizon computation requires low \textit{bin interval} for improved time for computation, and is not constrained with infinitesimal value for the interval.
Large bin intervals upto around \texttt{90} degrees, could be used in horizon computation, and is possible by a merge of the boundary points set with maximal bin points set.
The time for the construction using horizon computation, is greater than contour scanning because, an additional $O(n)$, operation is required to find the data points above horizon angles, and a superset of the points on the contour is computed.

\end{document}